%% file: main.tex
\documentclass[final, 10pt, twocolumn, conference]{IEEEtran}
\usepackage{placeins}
\usepackage{setspace}
\usepackage{dirtytalk}

\usepackage{amsmath}
\usepackage{amsthm}
\usepackage{amssymb}
\usepackage{enumitem}
\setlist[enumerate]{topsep=0pt,itemsep=-1ex,partopsep=1ex,parsep=1ex,leftmargin=4ex}
\setlist[itemize]{topsep=0pt,itemsep=-1ex,partopsep=1ex,parsep=1ex,leftmargin=4ex}
\usepackage{algorithm}
\usepackage{cite}
\usepackage{graphicx}
\usepackage{epstopdf}
\usepackage{url}
\usepackage{cite}
\usepackage{tabu}
\usepackage{texdef2015}
\usepackage{color}
\usepackage{booktabs} % For formal tables
\usepackage{xspace}
\usepackage{siunitx}
\usepackage{changes}
\usepackage{float}
\usepackage{balance}
\usepackage{multirow}
\usepackage{color, colortbl}
\usepackage{amssymb}
\usepackage{subfigure}
\usepackage{svg}
\allowdisplaybreaks
\usepackage{comment}

\usepackage{tikz}
\usetikzlibrary{automata,arrows,positioning,calc,fit,shapes.multipart,chains,shapes}
\usepackage{microtype}

\newcommand{\comm}[1]{}

\definecolor{LightCyan}{rgb}{0.88,1,1}
\IEEEoverridecommandlockouts
\begin{document}

%\title{Monte Carlo Evaluation of Dynamic Spectrum Allocation Techniques for Bandwidth Optimization in Wireless Communication Systems}
\title{Evaluation of Spectrum Sharing Algorithms for Networks with Heterogeneous Wireless Devices}

 % \author{\IEEEauthorblockN{
 %     Ankit Walishetti\IEEEauthorrefmark{1}, Igor Kadota\IEEEauthorrefmark{2}, Aidan Kim\IEEEauthorrefmark{3}, Colin Ward\IEEEauthorrefmark{4}, Eduardo Gutierrez\IEEEauthorrefmark{4}, Randall Berry\IEEEauthorrefmark{2}}
 % 	\IEEEauthorblockA{ \IEEEauthorrefmark{1}\small Illinois Mathematics and Science Academy, USA \IEEEauthorrefmark{2}\small Northwestern University, USA \\ \IEEEauthorrefmark{3}\small Rice University, USA \IEEEauthorrefmark{4}\small University of Illinois at Urbana-Champaign, USA\\
 % 	\IEEEauthorrefmark{1}\{awalishetti\}@imsa.edu \IEEEauthorrefmark{2}\{kadota, rberry\}@northwestern.edu \IEEEauthorrefmark{3}\{ak230\}@rice.edu \IEEEauthorrefmark{4}\{colinw3, eguti44\}@illinois.edu}}

    \author{Ankit Walishetti, Igor Kadota, Aidan Kim, Colin Ward, Eduardo Gutierrez, and Randall Berry% <-this % stops a space
\IEEEcompsocitemizethanks{\IEEEcompsocthanksitem 
A.\ Walishetti is with the Illinois Mathematics and Science Academy, USA. 
I.\ Kadota and R.\ Berry are with Northwestern University, USA. 
A.\ Kim is with Rice University, USA. 
C.\ Ward and E.\ Gutierrez are with University of Illinois at Urbana-Champaign, USA. 
This work was supported in part by the NSF under grant SES-2332054. A.\ Walishetti, A.\ Kim, C.\ Ward, and E.\ Gutierrez's research was supported by the Student Inquiry and Research (SIR) program from the Illinois Mathematics and Science Academy (IMSA). 
E-mail: awalishetti@imsa.edu, \{kadota, rberry\}@northwestern.edu, ak230@rice.edu, \{colinw3, eguti44\}@illinois.edu.}% <-this % stops an unwanted space
}

\maketitle
%\thispagestyle{plain}
%\pagestyle{plain}

%\vspace{-5mm}

\begin{abstract}
As highlighted in the National Spectrum Strategy, Dynamic Spectrum Access (DSA) is key for enabling 6G networks to meet the increasing demand for spectrum from various, heterogeneous emerging applications. In this paper, we consider heterogeneous wireless networks with multiple 6G base stations (BS) and a limited number of frequency bands available for transmission. Each BS is associated with a geographical location, a coverage area, and a bandwidth requirement. We assume that clients/UEs are within the corresponding BS's coverage area. To avoid interference, we impose that BSs with overlapping coverage areas must use different frequency bands. We address the challenging problem of efficiently allocating contiguous frequency bands to BSs while avoiding interference. Specifically, we define performance metrics that capture the feasibility of the frequency allocation task, the number of BSs that can be allocated within the limited frequency bands, and the amount of resources utilized by the network. Then, we consider five different DSA algorithms that prioritize BSs based on different features -- one of these algorithms is known in the graph theory literature as Welsh-Powell graph colouring algorithm -- and compare their performance using extensive simulations. Our results show that DSA algorithms that attempt to maximize the chances of obtaining a feasible frequency allocation -- which have been widely studied in the literature -- tend to under-perform in all other metrics. 
\end{abstract}

\section{Introduction}\label{sec:intro}
\input{src/intro}

\section{Problem Formulation}\label{sec:network_overview}
\input{src/background}

\section{Sorting Algorithms}\label{sec:sorting_methods}
% \subsection{AA}
% \subsubsection{BB}
\input{src/dsa}
\section{Simulation Results}\label{sec:sim_results}
\input{src/simulation_part2}

\section{Final Remarks}\label{sec:conclusion}
\input{src/conclusion}

% \section{Acknowledgements}
%This work was supported in part by NSF grants CNS-1827923, CNS-1836901, CNS-1910757, CNS-2148128, OAC-2029295, and AST-2037845. %\igor{Perhaps we can check consistency of references and add a few references, especially from Globecom or from Globecom TPC members?}
%\vspace{-3mm}
% This work was supported in part by NSF grants CNS-2148128, CNS-1827923, OAC-2029295, AST-2232455, AST-2232456, AST-2232459, EEC-2133516

\bibliographystyle{IEEEtran}
\bibliography{src/ref.bib}

\end{document}

%% file: src/intro.tex
% In recent years, the heightening demand for spectrum and bandwidth from wireless communication services has necessitated the development of more effective communication methods and technologies. Existing research underscores the necessity of these developments for next-generation applications including augmented reality, smart-cities, and the Internet of Things~\cite{ahmad20205g}. The ubiquity of cellular devices also drives this demand. As the proliferation of cellular devices and the complexity of their communications increases, the use of radio waves for communications increases correspondingly. 

Emerging and future applications and wireless devices will increasingly rely on Dynamic Spectrum Access (DSA) algorithms that can manage the scarce spectrum resources efficiently. 
The importance of DSA for 6G networks was recently highlighted in the National Spectrum Strategy~\cite{nationalSpectrum} and its Implementation Plan~\cite{nationalSpectrumPlan}, as well as in reports by the Next G Alliance~\cite{NGA}, Qualcomm~\cite{qualcomm}, Ericsson~\cite{ericsson}, and many others. 
The development of DSA algorithms that can efficiently allocate frequency spectrum to wireless devices (e.g., 6G base stations) while avoiding harmful interference has been extensively investigated in the literature (see surveys~\cite{survey_1,survey_2,survey_3}). 
\emph{Spectrum sharing is a challenging problem even in simple settings}. 
For example, consider \emph{homogeneous networks} in which every base station has the same bandwidth requirement. The problem of finding the minimum number of frequency bands that can accommodate multiple homogeneous base stations is known to be NP-hard~\cite{GAREY1976237}. 
Several heuristic DSA algorithms for homogeneous networks have been proposed in the graph theory literature~\cite{greedy_coloring,DSATUR,DSATUR_2,vtc,ICNC,Kadota_WiNTECH}. 
Heterogeneous networks, in which base stations may have diverse bandwidth requirements, are more complex and less studied. Recent focus has been on developing machine learning-based DSA algorithms for heterogeneous networks, e.g.~\cite{rl_1,rl_2,rl_3}. 

%The growing ubiquity of Internet of Things (IoT) devices paired with the rise of high-bandwidth applications places constant pressure on existing wireless networks to deliver reliable and high-speed connectivity. As the proliferation of bandwidth-hungry devices and the complexity of their communications increases, the use of radio waves for communications intensifies correspondingly. The heightening demand for spectrum and bandwidth space from wireless communication services has necessitated the development of more effective spectrum management methods. Existing research underscores the necessity of these developments for next-generation applications including augmented reality, smart-cities, and the Internet of Things (Netalkar et al., 2023). Namely, Dynamic Spectrum Allocation (DSA) shows promise in the way it allows for the flexible mapping of transmitter devices in heterogeneous networks. DSA technology is capable of assessing channel occupancy across spectrum space before performing device allocation, making it ideal for exploiting spectrum sharing opportunities []. This cognitive radio-based spectrum switching technology is powered by spectrum sensing, which involves obtaining the spectrum usage characteristics []. In this way, features of licensed transmitter devices are extracted to understand limitations of allocation, such as with geographic overlap. 

%Prior works, such as [], utilize DSA to detect idle, unlicensed bands in an effort to reduce waste of spectrum resources by maximizing the utilization of the spectrum.  
In this paper, we evaluate different low-complexity DSA algorithms for \emph{heterogeneous} 6G networks, and use extensive simulation results to gain insight into their performance trade-offs. 
We define performance metrics that capture the feasibility of the frequency allocation task, the number of base stations that can be allocated within the limited frequency bands, and the amount of resources utilized by the network. 
Then, we compare DSA algorithms that prioritize frequency allocation to base stations based on different features, including their number of potentially interfering neighbors, their coverage area, and their required bandwidth. 
Our simulation results show that DSA algorithms that prioritize frequency allocation to base stations with higher number of potentially interfering neighbors are more likely to achieve a feasible frequency allocation (for all base stations in the network) within the limited available frequency. This result agrees with the literature on homogeneous networks~\cite{greedy_coloring}. However, our results also show that, when the frequency allocation task is unfeasible, these DSA algorithms are more likely to leave a high number of unallocated base stations. Performance trade-offs associated with other DSA algorithms are discussed in Section~\ref{sec:sim_results}.

The rest of this paper is organized as follows. 
Section~\ref{sec:network_overview} formulates the DSA problem. %introduces the network model, the DSA algorithm, and the performance metrics.
In Section~\ref{sec:sorting_methods}, we develop five distinct sorting algorithms. %that embrace spectrum sensing and sharing opportunities for DSA. 
In Section~\ref{sec:sim_results}, we describe the simulation platform and discuss our extensive numerical results. %design a DSA strategy for efficient utilization of spectrum space.
%Section~\ref{sim_results} describes the simulation platform developed to evaluate our sorting algorithms across our range of performance metrics.  
%Section 6 discusses the simulation results.
%Section~\ref{simulation} describes the simulation platform developed to evaluate SCM-based DSA algorithms and discusses simulation results for the proposed algorithm. %Section~\ref{simulation} describes our DSA algorithm along with a simulation platform we designed to evaluate multiple DSA schemes using SCMs. 
Section~\ref{sec:conclusion} concludes the paper. 
%Section~\ref{future} provides directions for future experimentation and Section~\ref{conclusion} concludes the paper.

%First, we present the results of our experimental work on the initial design of DSA algorithms that leverage the capabilities of SCM

%    \item \textit{\textbf{Intermodulation mask:}} Defines the propensity of co-located signals to generate intermodulation products in a transmitter or receiver.
%    \item \textit{\textbf{Platform Name:}} A name or list of names of platforms that are attributed to a particular location/platform (i.e. airplane, vehicle, etc.). They are useful in identifying when multiple systems are co-located and could suffer intermodulation and out-of-band interference effects.  
%    \item \textit{\textbf{Minimum power spectral flux density:}} A power spectral flux density that when used as part of a transmitter model implies the geographical extent in which receivers in the system are protected.
%    \item \textit{\textbf{Policy or protocol:}} A named protocol or policy with parameters that define behaviors supported by an RF device or system that allow different systems to be co-located and to co-exist in the same spectrum.    

%% file: src/background.tex
% SCMs use a set of 11 data elements, referred to as constructs, to describe the spectral, spatial, and temporal characteristics of spectrum use by any RF device and/or system. These constructs, which are defined in the IEEE 1900.5.2 standard~\cite{scmstdofficial,scmstandard}, can be used to build different types of SCMs, including:
% %\begin{itemize}[leftmargin=*]
%     (i) \textit{\textbf{Transmitter models}} that convey the extent and strength of RF emissions from a transmitter; 
%     (ii) \textit{\textbf{Receiver models}} that convey what harmful interference to an RF receiver device is; and 
%     (iii) \textit{\textbf{System and Set models}} that group several transmitter and receiver SCMs. 
% %\end{itemize}
% % The SCM constructs most relevant to this work are described below. %We will focus on the mandatory constructs. 
% Unless otherwise stated, the constructs must be used in both transmitter and receiver models. 
%\subsection{Network Model}\label{}
\begin{figure}[t]
%\vspace{-5mm}
\centering
\centerline{\includegraphics[width=0.50\columnwidth]{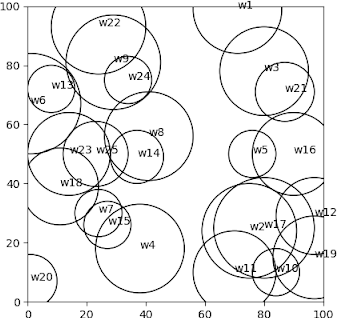}}
\vspace{-3mm}
\caption{Illustration of $N=25$ transmitters located in a two-dimensional region $\mathcal R$ of  $100\times100$ square meters. Each transmitter is associated with a coverage area. To prevent harmful interference, it is imposed that transmitters with overlapping coverage cannot use the same frequency bands.}
\label{fig:result_map}
\vspace{-3mm}
\end{figure}

\begin{figure}[t]
%\vspace{-4mm}
\centering
\centerline{\includegraphics[width=0.65\columnwidth]{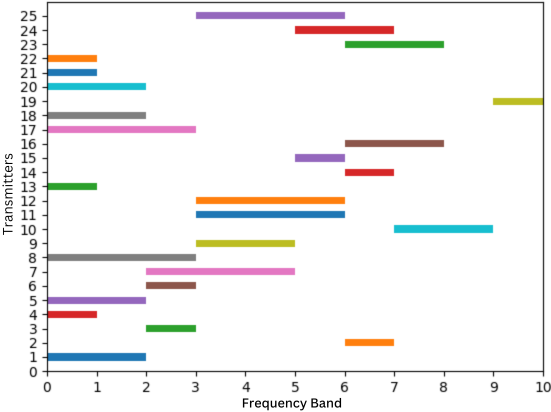}}
\vspace{-3mm}
\caption{%Outcome of a DSA algorithm that assigns a frequency bandwidth of $B_i\in\{1,2,3\}$ units to each of the $N=25$ transmitters in Figure~\ref{fig:result_map} while preventing harmful interference.
Outcome of a DSA algorithm for the $N=25$ transmitters in Figure~\ref{fig:result_map} and a total of $F=10$ units of available bandwidth. 
Each transmitter is allocated its required bandwidth $B_i$ while avoiding interference. 
Transmitter~``3'' (identified as ``w3'' in Figure~\ref{fig:result_map}) overlaps with ``1'' and ``21.'' 
Transmitters ``1'' and ``21,'' which do not overlap, are assigned bandwidths $f\in\{1,2\}$ and $f\in\{1\}$, respectively, while ``3'' is assigned $f\in\{3\}$ to avoid interference. Transmitter ``19'' overlaps with ``10,'' ``11,'' ``2,'' ``17,'' and ``12.'' The only bandwidth available for transmitter ``19'' is $f\in\{10\}$.}
\label{fig:result_dsa_scheme}
\vspace{-6mm}
\end{figure}

In this section, we discuss the network model, the DSA algorithm, and the performance metrics. %employed to evaluate the algorithm. 

\vspace{0.5ex}
\noindent\textbf{Network Model.} We consider a network with multiple 6G base stations, called transmitters.\footnote{Though we refer to these as transmitters, our model can apply to either the uplink or the downlink in a Frequency Division Duplex (FDD) deployment or to a Time Division Duplex (TDD) deployment in which each base station operates as both a transmitter and receiver.} 
Each transmitter is associated with a (potentially) different geographical location, coverage area, and frequency bandwidth. 
Let $N$ be the total number of (heterogeneous) transmitters in the network, and let $F$ be the total number of (contiguous) bandwidth units available for transmission.
Each transmitter $i\in\{1,2\ldots,N\}$ is associated with a location within a two-dimensional region $\mathcal R$, a bandwidth requirement of $B_i$ units, and a transmission coverage of $R_i$ meters. We assume that there is a single band manager for this region that collects transmitter requirements and runs a DSA algorithm to allocate bandwidth. 
%We assume that the total bandwidth $F$ and the bandwidth requirements $B_i$ are contiguous.
%Our network model represents a heterogeneous environment with varying transmitter characteristics including transmitter coverage, bandwidth requirements, and geographical distribution. We identify transmitters on a two-dimensional geographic map using two randomly generated parameters: a coordinate pair denoting their node and signal propagation radii delineating the extent of their coverage range. 
We model the coverage of transmitter $i$ as a circle with radius $R_i$ centered at the location of the transmitter, as can be seen in Figure~\ref{fig:result_map}.\footnote{This models a scenario where transmitters have omni-directional antennas. Most of the following can be extended to allow for more general coverage regions.} 
We assume that clients (also known as User Equipment) are within the corresponding transmitter's coverage area.\footnote{Depending on the requirements of the users, the user deployment could be constrained to be sufficiently inside the coverage area to allow for "guard regions" between different deployments to reduce interference.} 
Hence, if two (or more) transmitters' coverage areas overlap, they may cause harmful interference to one another. 
\emph{To avoid interference, we impose that transmitters with overlapping coverage areas must use different frequency bands.} 
%We model each transmission area as a circle with a given radius, forming the basis of a unit disc graph [cite?] as seen in Figure 1. 
%Given a random setup of transmitters, the overlapping coverage of each transmitter must be taken into account. We consider a scenario where transmitters with overlapping coverage areas, or interference, in geographical spaces must be assigned distinct frequencies inside spectrum space to avoid risking signal conflict issues and ensure thorough data transmission to each transmitter in the system. However, 
Naturally, non-overlapping transmitters may use the same frequency bandwidth, %playing into the idea of opportunistic spectrum usage and 
allowing for more efficient spectrum utilization. %In this network model, transmitters represent 6G base stations and receivers represent clients associated with the base stations. 
%In our network model, the task of transmitter assignment is managed by both a sorting algorithm and an allocation algorithm, as we explain later in Section 2. 
The effects of aggregate interference and time-varying coverage are not considered. 

% \subsection{2.2 Parameterization}\label{}
% The design structure of our network model involves several key parameters that dictate the network structure and affect the performance of the system. Within their respective ranges, parameters such as the available bandwidth, geographical area, transmitter coverage, and the number of transmitters are changed individually to gauge their effect on performance in various scenarios. Note that all numbers are in arbitrary units. Geographical area ranges from 50$\times$50 units to 170$\times$170 units, with x-values and y-values increasing with every integer value. Transmitter bandwidth requirements range from 1 to 3, inclusive of all integer values ($1 \leq B_{T_i} \leq 3$). Transmitter coverage varies from 8 to 17 ($8 \leq R_{T_i} \leq 17$). The number of transmitters ranges from 5 to 30 $\{ T_x \mid 5 \leq x \leq 30 \}$ and the available bandwidth intervals starts at 10 and increases as needed ($B_i \geq 10$ intervals).

%\subsection{2.3 Transmitter Ordering and Allocation}\label{}
\vspace{0.5ex}
\noindent\textbf{Dynamic Spectrum Access algorithm.} The DSA algorithm has two main components: a sorting algorithm and a frequency allocation mechanism. 
The \emph{sorting algorithm} takes the list of transmitters $(1,2,\ldots,N)$ and re-arranges it in a particular order $(i_1,i_2,\ldots,i_N)$. 
Different sorting algorithms prioritize transmitters based on different performance metrics (introduced below). 
The \emph{frequency allocation mechanism} assigns a contiguous bandwidth to each transmitter, one at a time, following the order $(i_1,i_2,\ldots,i_N)$. 
Specifically, for each transmitter $i$, the frequency allocation mechanism assigns the $B_i$ units of available bandwidth with the lowest possible indices $f\in\{1,2,\ldots,F\}$.\footnote{We assume that the bandwidth assigned to a transmitter must be contiguous. This can help to minimize the amount of spectrum needed for guard-bands, but does reduce flexibility in making assignments. We leave the issue of non-contiguous bandwidth for future work.} Recall that transmitter $i$ may overlap with other transmitters, which may restrict its available bandwidth. 
Intuitively, the frequency allocation mechanism is attempting to pack transmitter bandwidths as much as possible, leaving contiguous available bandwidth for other transmitters in the network. 
The goal of the \emph{Dynamic Spectrum Access algorithm} is to assign frequency bandwidths to all $N$ transmitters while avoiding harmful interference. The computational complexity of these DSA algorithms is dominated by the sorting algorithm which has complexity $O(N\log(N))$ when using the Heap Sort approach. 
In Sec.~\ref{sec:sorting_methods}, we introduce different sorting algorithms and, in Sec.~\ref{sec:sim_results}, we evaluate the impact of these different sorting algorithms.

\vspace{0.5ex}
\noindent\textbf{Performance Metrics.} To evaluate DSA algorithms based on different sorting algorithms, we employ the five metrics defined below. %First, we define performance metrics for networks with a fixed number of transmitters $N$. Then, we define performance metrics for networks with an increasing $N$. 
Given a DSA algorithm, a network with $N$ transmitters can be classified as \emph{feasible} or \emph{unfeasible}. The network is feasible if the DSA algorithm allocates frequency bands within the range $f\in\{1,2,\ldots,F\}$ for every transmitter $i\in\{1,2,\ldots,N\}$. In this case, all transmitters are classified as \emph{admissible}. The network is unfeasible if there is at least one \emph{inadmissible} transmitter that is allocated frequency bands outside of the range, i.e., $f\geq F+1$. Denote by $\mathcal{N}$ the set of admissible transmitters and by $\mathcal{N}'$ the set of inadmissible transmitters. It follows that $\mathcal{N}\cap\mathcal{N}'=\emptyset$ and $\mathcal{N}\cup\mathcal{N}'=\{1,2,\ldots,N\}$. 
%The sorting algorithms discussed in Sec.~\ref{sec:sorting_methods} are designed to optimize one or more of these metrics. 
%To ensure our network model achieves desired performance goals, it is crucial to define and utilize objective functions. These functions guide the sorting and allocation of transmitters, helping us quantify and optimize performance. Given the diverse array of transmitter characteristics and parameters, there exists a multitude of ways to measure the performance of a sorting algorithm. The method by which these transmitters are allocated in the spectrum varies the bandwidth used and other outcomes greatly (Netalkar et al., 2023). To investigate the efficacy and robustness of our allocation strategies we present 7 \emph{performance metrics} that align with real-world measures of efficiency: 
%
%Performance metrics of a DSA algorithm for a network with a fixed number of transmitters $N$:
\begin{itemize}[leftmargin=*]
    \item \underline{Feasibility Indicator} ($FI\in\{0,1\}$) is a binary metric that indicates whether the network is feasible ($FI=1$) or unfeasible ($FI=0$). 
        \item \underline{Bandwidth Usage} ($BU\in\mathbb{N}^+$) represents the highest bandwidth index $f\in\{1,2,\ldots,F,F+1,\ldots\}$ allocated by the DSA algorithm to any of the transmitters $i\in\{1,2,\ldots,N\}$. In Figure~\ref{fig:result_dsa_scheme}, we have $BU=F=10$. If the network is feasible ($FI=1$), then $BU\leq F$. If the network is unfeasible ($FI=0$), then $BU\geq F+1$. A low $BU$ indicates that the DSA algorithm can effectively pack transmitters within the limited bandwidth. % $f\in\{1,2,\ldots,F\}$. 
    \item \underline{Coverage Area} ($CA\in\mathbb{R}^+$) represents the sum of the coverage areas of every \emph{admissible} transmitter
    \begin{equation}
        CA = \textstyle\sum_{i\in\mathcal{N}} \pi R_i^2 C_{i} \; ,
    \end{equation}
    where $C_{i}\in[0,1]$ is a correction factor accounting for the fraction of the coverage area within the two-dimensional region $\mathcal R$. For instance, ``w1'' in Figure~\ref{fig:result_map} has a $C_{1}\approx 0.5$. 
    \item \underline{Bandwidth-Coverage Product} ($BC\in\mathbb{N}^+$) is the sum of the product of the coverage radius and the bandwidth requirement of every admissible transmitter\footnote{If the coverage regions were not circles, one can define an ``effective radius'' for each region and still use this algorithm.  Alternatively, one could also consider a related metric given by the product of the bandwidth and the coverage area.}
    \begin{equation}
         BC = \textstyle\sum_{i\in\mathcal{N}} R_i B_i \; .
    \end{equation}
    A high $R_i B_i$ indicates a transmitter with stringent bandwidth requirement $B_i$ and large coverage radius $R_i$, i.e., a transmitter that consumes a significant amount of resources. A high $R_i B_i$ may also indicate an important transmitter that serves many clients and can support high quality of service. %... TBD. A transmitter with a high bandwidth requirement $B_i$ and a high coverage radius $R_i$ is expected to deliver... %high quality of service to where \( R_{ij} \) is the coverage radius of the \( i \)-th transmitter in the \( j \)-th trial, \( B_{ij} \) is the bandwidth requirement of the \( i \)-th transmitter in the \( j \)-th trial, \( m \) is the number of transmitters, and \( n \) is the number of trials.
    \item \underline{Total Transmitters while Feasible} ($TF\in\mathbb{N}^+$) represents the maximum number of transmitters before the network becomes unfeasible. Recall that the sorting algorithm takes the list of transmitters $(1,2,\ldots,N)$ and sorts it in a particular sequence $(i_1,i_2,\ldots,i_N)$. If the network is feasible, then all $N$ transmitters are admissible and Total Transmitters while Feasible is $TF=N$. Otherwise, let $N'$ be the sorted index of the first inadmissible transmitter. It follows that transmitters in $(i_1,i_2,\ldots,i_{N'-1})$ are all admissible, transmitter $i_{N'}$ is inadmissible, and transmitters in $(i_{N'+1},i_{N'+2},\ldots,i_N)$ may be admissible or inadmissible. In this case, $TF=N'-1$. 
\end{itemize}
These metrics provide a comprehensive framework for evaluating the performance of different DSA algorithms. %our transmitter ordering and allocation strategies, helping us understand which ones align with real-world requirements and cater towards the optimal DSA system. 
For instance, Feasibility Indicator, $FI$, and Bandwidth Usage, $BU$, have direct connections with the system objective and resource allocation efficiency. Measuring the Total Transmitters while Feasible, $TF$, helps us understand how close/far the DSA algorithm was to achieving network feasibility. %\emph{Feasibility} is a direct measure of a network system’s performance while the \emph{bandwidth usage} metric correlates with the resource-management capabilities of the system. Ultimately, these metrics direct us in refining our approach to achieve a robust and efficient DSA system.

%% file: src/dsa.tex
% \subsection{Compatibility test for a transmitter-receiver pair}\label{compatibility_test}
% The Compatibility Test (CT) using SCMs aims at determining if a transmitter model is compatible with a receiver model. CT begins by checking if the SCMs overlap in both time and frequency. If they do not overlap, there is compatibility. If they do overlap, then the evaluation continues. 

In this section, we develop five sorting algorithms %to dictate the order of allocation for a given list of transmitter devices. Each algorithm follows a unique DSA-based approach, 
that prioritize transmitters according to different characteristics. The combination of each sorting algorithm with the frequency allocation mechanism described in Sec.~\ref{sec:network_overview} composes a different DSA algorithm. We will use numerical results to evaluate these different DSA algorithms in Sec.~\ref{sec:sim_results}. The description of the sorting algorithms is below:
%leverages these sorting algorithms and the frequency allocation mechanism described in Sec.~\ref{sec:network_overview} to optimize the frequency assignment to the $N$ transmitters in the network while preventing harmful interference. 
%Together with the frequency allocation mechanism described in Sec.~\ref{sec:network_overview}, these sorting algorithms aiming to optimize frequency allocation while preventing harmful interference. %For example, when considering transmitter characteristics such as signal propagation range and bandwidth requirement, a tradeoff between prioritizing many smaller transmitters or a few prominent transmitters is studied.  
%
% \begin{table}[!h]
% \centering
% \small
% \vspace{-2mm}
% \caption{Sorting Algorithms}
% \begin{tabular}{lp{5cm}}
% \toprule
% \textbf{Sorting Algorithm} & \textbf{Characteristic of Priority}\\ \midrule
% Random Sort & Set of randomly allocated transmitters\\ 
% Bandwidth-Coverage Product Sort & Transmitter power (radius*bandwidth)\\ 
% Most-Overlaps Sort & Highest overlap (interference)\\ 
% Least-Bandwidth Sort & Smallest bandwidth requirements\\ 
% Least-Coverage Sort & Smallest coverage\\ 
% \bottomrule
% \end{tabular}
% \label{tab:sim_parameters}
% \vspace{-2mm}
% \end{table}
\begin{itemize}[leftmargin=*]
    \item \underline{Most-Overlaps Sort} prioritizes transmitters in descending order\footnote{Ties in every sorting algorithm are broken arbitrarily. For example, by prioritizing transmitters with lowest index $i\in\{1,2,\ldots,N\}$} of number of coverage overlaps. The number of coverage overlaps associated with transmitter~$i$ represents the number of neighboring transmitters that can cause harmful interference to transmitter~$i$. The number of overlaps is equivalent to the degree of the conflict graph~\cite{conflict_graph} underlying the network under consideration. In the context of graph theory, the sorting algorithm that prioritizes transmitters in descending order of degree is called the Welsh-Powell graph colouring algorithm~\cite{greedy_coloring} and is widely used (together with its variants, including the DSATUR algorithm~\cite{DSATUR,DSATUR_2}) to find the minimum number of frequency bands $F$ needed to make a network feasible. The main idea of this algorithm is to start the frequency allocation process with the most challenging transmitters. 
    \item \underline{Bandwidth-Coverage Sort} prioritizes transmitters in descending order of bandwidth-coverage product $R_i B_i$. Intuitively, this prioritizes transmitters that need the most resources. Transmitters with high $R_i B_i$ may be considered more important, as they may be able to serve more clients and/or provide superior quality of service. 
    \item \underline{Least-Bandwidth Sort} prioritizes transmitters in ascending order of bandwidth requirement $B_i$. This algorithm contrasts with the previous algorithms in that it starts the frequency allocation process with the least challenging transmitters. 
    \item \underline{Least-Coverage Sort} prioritizes transmitters in ascending order of coverage radius $R_i$. %By comparing with Bandwidth-Coverage Sort and the well-known Most-Overlaps Sort, we will hig
    \item \underline{Random Sort} arranges the $N$ transmitters randomly. Random sort serves as a benchmark for comparison with other sorting algorithms. 
\end{itemize}

These sorting algorithms provide diverse strategies for transmitter allocation. Next, we simulate and compare these strategies, highlighting their performance tradeoffs %each with its specific objectives and theoretical tradeoffs. With a comparative analysis of the developed methods, we aim to identify the most effective DSA approaches for optimizing frequency allocation and minimizing interference 
in various heterogeneous network scenarios. Recall that the computational complexity of these algorithms is $O(N \log(N))$. 

%% file: src/simulation_part2.tex
% To evaluate the performance of our proposed SD algorithm, we use a Python-based simulator that assigns Tx/Rx pairs to random locations on a given operational area and then uses Algorithm \ref{algo:1} to deconflict spectrum use. The simulation uses an event-based framework with Tx/Rx pairs joining the system sequentially, i.e., one after the other. The simulation parameters are summarized in Table~\ref{tab:sim_parameters}. To perform CTs, we leverage Octave code from the Spectrum Consumption Model Builder and Analysis Tool (SCMBAT)~\cite{SCMBAT_website} which interfaces with our python simulator using \textit{oct2py}. 

In this section, we evaluate five DSA algorithms in terms of the performance metrics described in Sec.~\ref{sec:network_overview}, namely the Feasibility Indicator, $FI$, the Bandwidth Usage, $BU$, the Coverage Area, $CA$, and the Bandwidth-Coverage Product, $BC$. Each DSA algorithm is a combination of the frequency allocation mechanism described in Sec.~\ref{sec:network_overview} with one of the five sorting algorithms developed in Sec.~\ref{sec:sorting_methods}, namely Bandwidth-Coverage Sort, Most-Overlaps Sort, Least-Bandwidth Sort, Least-Coverage Sort, and Random Sort. The DSA algorithm is named after the sorting algorithm. 

\begin{figure*}[t]
  \centering
  \subfigure[]{\includegraphics[width=0.32\textwidth]{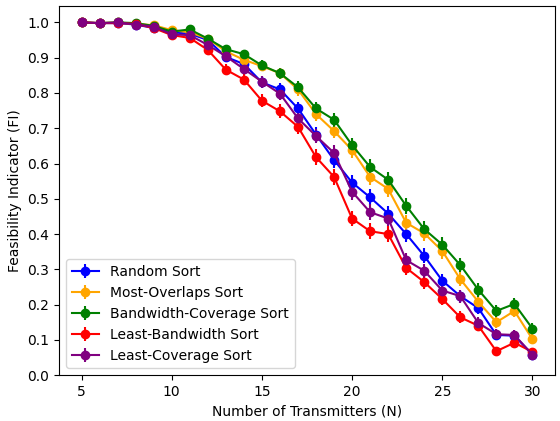}}
  \hspace{0.5cm}
  \subfigure[]{\includegraphics[width=0.32\textwidth]{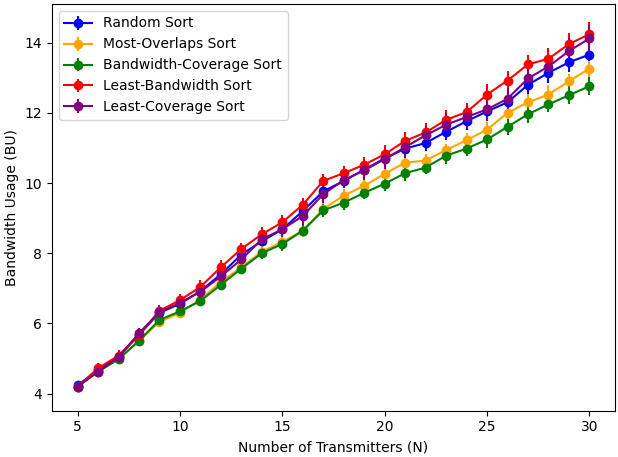}}    
  \hspace{0.5cm}
  \subfigure[]{\includegraphics[width=0.32\textwidth]{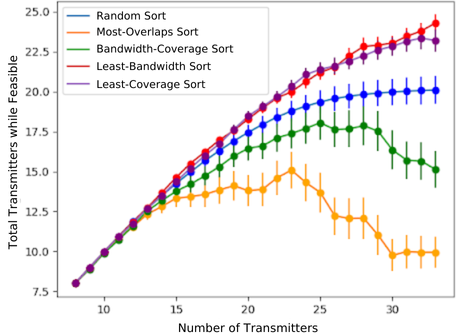}}
  \hspace{0.5cm}
  \subfigure[]{\includegraphics[width=0.32\textwidth]{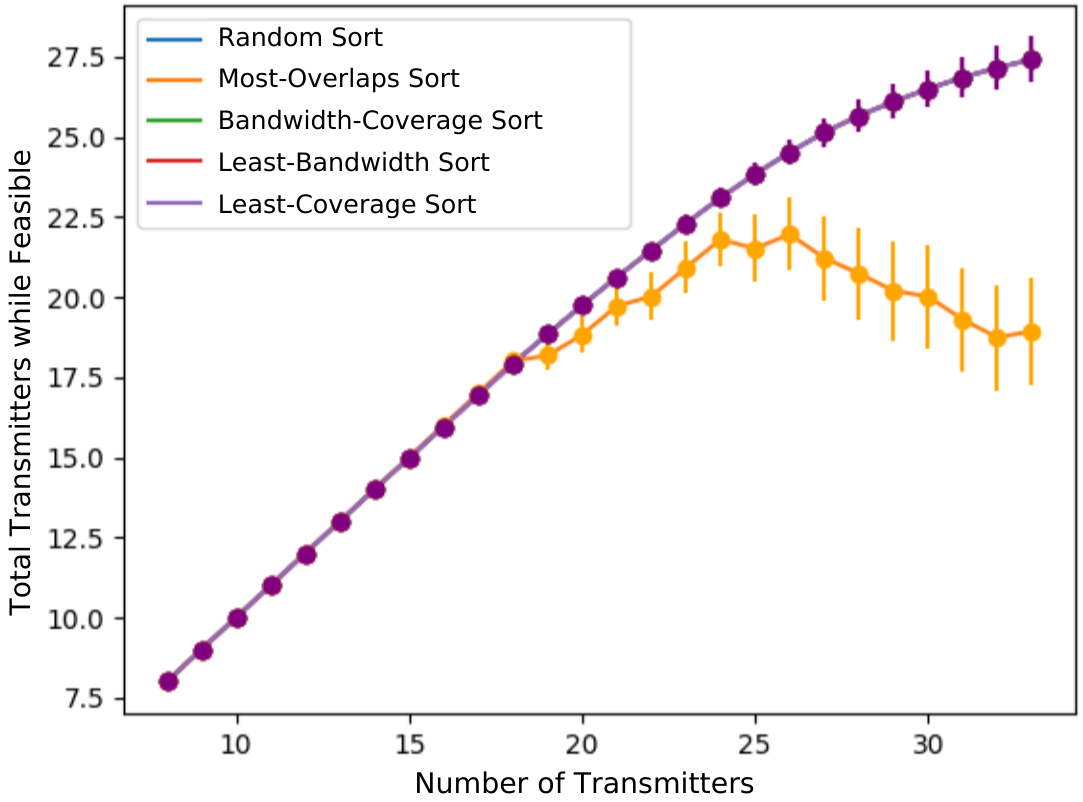}}
  \vspace{-0.3cm}
  \caption{Performance of networks with increasing number of transmitters $N\in\{5,6,\ldots,30\}$. Unless indicated otherwise, network parameters are set according to Table~\ref{tab:sim_parameters}. (a), (b), and (c) represent heterogeneous networks with coverage $R_i \sim Uniform(8, 17)$ and bandwidth requirement $B_i \sim Uniform(1, 3)$. (d) represents homogeneous networks with $R_i = 12,\forall i,$ and $B_i = 2,\forall i$. (a) shows the Feasibility Indicator, $FI$, averaged over $500$ simulation runs. (b) shows the Bandwidth Usage, $BU$, averaged over $50$ simulation runs. (c) and (d) show the Total Transmitters while Feasible, $TF$, averaged over $50$ simulation runs. Notice that the best performing DSA algorithms in (a) and (b) are the worse in (c).}\label{fig:simulation_increasing_N}
  \vspace{-0.4cm}
\end{figure*}

\begin{figure*}[t]
  \centering
  \subfigure[]{\includegraphics[width=0.34\textwidth]{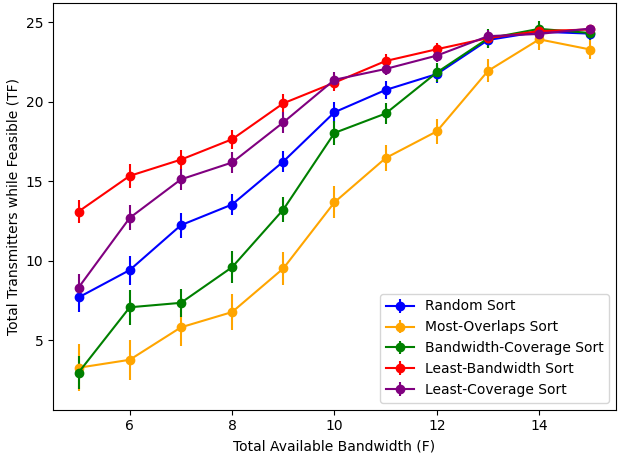}}
  %\subfigure[]{\includegraphics[width=0.33\textwidth]{images/TF_F_final.png}}
  \hspace{0.5cm}
  \subfigure[]{\includegraphics[width=0.34\textwidth]{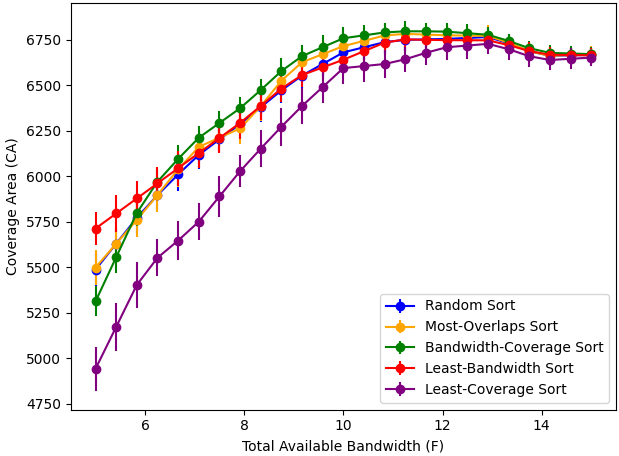}}
  %\hspace{0.5cm}
  \vspace{-0.3cm}
  \caption{Performance of networks with increasing total available bandwidth $F\in\{5,6,\ldots,15\}$. (a) shows the Total Transmitters while Feasible, $TF$. (b) shows the Coverage Area, $CA$.}\label{fig:simulation_increasing_F}
  \vspace{-0.4cm}
\end{figure*}

\begin{figure*}[t]
  \centering
  \subfigure[]{\includegraphics[width=0.34\textwidth]{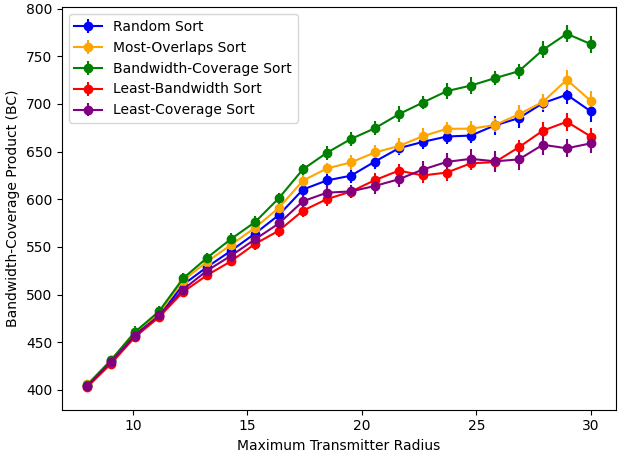}}
  %\subfigure[]{\includegraphics[width=0.33\textwidth]{images/TF_F_final.png}}
  \hspace{0.5cm}
  \subfigure[]{\includegraphics[width=0.34\textwidth]{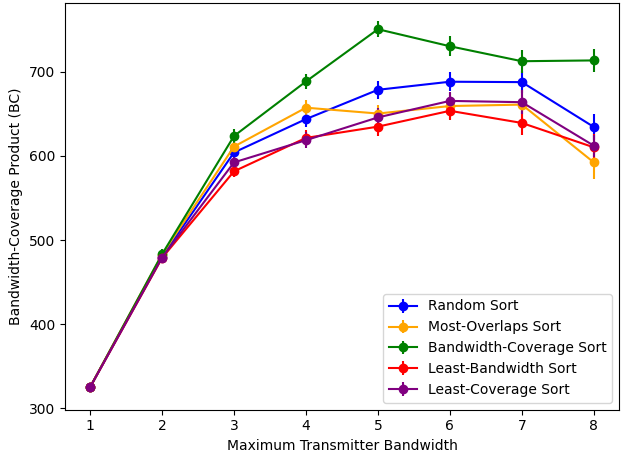}}
  %\hspace{0.5cm}
  \vspace{-0.3cm}
  \caption{Performance of networks in terms of the Bandwidth-Coverage Product, $BC$, associated with their admissible transmitters. A high $BC$ indicates a DSA algorithm that can allocate transmitters that can cover a large area and can support high quality of service. (a) and (b) represents networks with increasing heterogeneity. (a) has $N=25$ transmitters, each with coverage radius sampled according to $R_i \sim Uniform(8, R_{max})$ with increasing maximum transmitter radius $R_{max}\in\{8,9,\ldots,30\}$. (b) has $N=25$ transmitters, each with required bandwidth sampled according to $B_i \sim Uniform(1, B_{max})$ with increasing maximum transmitter bandwidth $B_{max}\in\{1,2,\ldots,8\}$. Other parameters are set as in Table \ref{tab:sim_parameters}.}\label{fig:simulation_increasing_R_B}
  \vspace{-0.4cm}
\end{figure*}

\vspace{0.5ex}
\noindent\textbf{Network Simulator.} To evaluate the DSA algorithms, we use an object-oriented, Python-based simulator based on Monte-Carlo methods. Specifically, given the total number of bandwidth units available for transmission $F$ and the total number of transmitters $N$, the simulator \emph{randomly} assigns a geographical location, a required bandwidth $B_i$, and a coverage radius $R_i$ for each transmitter $i\in\{1,\ldots,N\}$. The location is assigned uniformly at random within a two-dimensional region $\mathcal R$ defined as a $100 \times 100$ meters square. The required bandwidth $B_i$ and coverage radius $R_i$ are assigned uniformly at random from specific ranges. The ranges of $B_i$ and $R_i$ and the values of $N$ and $F$ are as specified in Table~\ref{tab:sim_parameters}, unless noted otherwise. After establishing the network setup, the simulator runs the five DSA algorithms and computes their associated performance metrics. To account for the effects of different network setups, for each set of network parameters, we create at least $50$ randomly generated networks setups and display in Figures~\ref{fig:simulation_increasing_N},~\ref{fig:simulation_increasing_F}, and~\ref{fig:simulation_increasing_R_B} the average of the metrics associated with these multiple simulation runs. The error bars in Figures~\ref{fig:simulation_increasing_N},~\ref{fig:simulation_increasing_F}, and~\ref{fig:simulation_increasing_R_B} represent the standard deviation.

\begin{table}[t]
\centering
\small
\vspace{-2mm}
\caption{Baseline Simulation Parameters}
\begin{tabular}{lc}%||l||l|
%\hline
%\hline
\toprule
\textbf{Parameter} & \textbf{Value} \\ \midrule %\hline
%Latitude (min, max)      & (40.805726, 40.812829)\\
%Longitude (min, max)      & (-73.945219, -73.955432)\\
Geographical Area&   $100\times100$ $m^2$\\ %\hline
Number of Transmitters&  $N=$ 25 \\ %\hline
Total Available Bandwidth&  $F=$ 10 units\\ %\hline
Bandwidth of Transmitter $i$&   $B_i \sim Uniform(1, 3)$\\ %\hline
Coverage of Transmitter $i$&   $R_i \sim Uniform(8, 17)$\\ %\hline
Number of Simulation Runs&  $50$\\ %\hline
\bottomrule
\end{tabular}
\label{tab:sim_parameters}
\vspace{-4mm}
\end{table}

\vspace{0.5ex}
\noindent\textbf{Numerical Results.} In Figures~\ref{fig:simulation_increasing_N}(a)-(d), we simulate networks with increasing number of transmitters $N\in\{5,6,$ $\ldots,30\}$ and display the performance of DSA algorithms in terms of the Feasibility Indicator, $FI$, Bandwidth Usage, $BU$, and the Total Transmitters while Feasible, $TF$. 
%looking to assess how the feasibility of networks is impacted as demand increases. This analysis enables a direct comparison between the five sorting algorithms, each represented on the same set of axes to allow for easier interpretation of their performance differences. Figure 3 indicates that 
Figures~\ref{fig:simulation_increasing_N}(a), (b), and (c) show $FI$, $BU$, and $TF$, respectively, associated with networks with the exact same parameters. 
Interestingly, the best-performing DSA algorithms in Figures~\ref{fig:simulation_increasing_N}(a) and (b) are the worse in Figure~\ref{fig:simulation_increasing_N}(c). 
%Figure~\ref{fig:simulation_increasing_N}(c) shows constrasting results to Figures \ref{fig:simulation_increasing_N}(a)and (b).
Figures~\ref{fig:simulation_increasing_N}(a) and (b) show that Most-Overlaps Sort and Bandwidth-Coverage Sort outperform Least-Bandwidth Sort and Least-Coverage Sort in terms of the average $FI$ and $BU$ \emph{in every simulation}, with the performance gap becoming more pronounced for networks with large number of transmitters $N>15$. %Most-Overlaps Sort and Bandwidth-Coverage Sort outperform %This observation suggests that the two sorting methods are particularly capable of meeting network demand. Figure 3 also points to a poor performance of 
%Least-Bandwidth Sort and Least-Coverage Sort in terms of the average Feasibility Indicator $FI$ in every simulation. %throughout the entire range of transmitter quantities. 
This result agrees with the graph theory literature~\cite{greedy_coloring,DSATUR,DSATUR_2} in that starting frequency allocation with the most challenging transmitters (e.g., the transmitters with most overlaps) is a promising approach for minimizing the bandwidth usage, thus improving the chances of feasibility. In the graph theory literature, the minimum bandwidth usage is called the chromatic number of the graph/network. %For this reason, Most-Overlaps Sort (which is equivalent to the Welsh-Powell graph colouring algorithm) is widely studied in the literature~\cite{greedy_coloring,DSATUR,DSATUR_2}. %As expected, Random Sort is neither the best or the worse sorting algorithm. 
%The results from Figure 3 align closely with theoretical expectations regarding higher degree and lower degree sorting methods.
%Figure~\ref{fig:simulation_increasing_N}(b) displays Bandwidth-Coverage Sort and Most-Overlaps Sort achieving lower levels of Bandwidth Usage $BU$ compared to Least-Bandwidth Sort and Least-Coverage Sort. Bandwidth-Coverage Sort's result can be explained by This advantage becomes more prominent for networks with large number of transmitters $N>15$. 

% \begin{figure}[h]
% \vspace{-2mm}
% \centering
% \centerline{\includegraphics[width=0.83\columnwidth]{images/Total_Transmitters_while_Feasible.png}}
% \vspace{0mm}
% \caption{Total Transmiters while Feasible $TF$ of networks with increasing transmitters quantities $N\in\{5,6,\ldots,30\}$ and other parameters set as defined in Table~\ref{tab:sim_parameters}. Results are averaged over $50$ simulation runs.}
% \label{fig:simulation_2}
% \vspace{-2mm}
% \end{figure}
%In Figure~\ref{fig:simulation_increasing_N}(b), we simulate networks with increasing transmitter quantities, focusing on measuring the number of transmitters assigned feasibly, that is, with $B_i \leq 10$. 

In contrast, Figure~\ref{fig:simulation_increasing_N}(c) shows that 
%Figure~\ref{fig:simulation_increasing_N}(b) indicates that 
Least-Coverage Sort and Least-Bandwidth Sort outperform Most-Overlaps Sort and Bandwidth-Coverage Sort in terms of the Total Transmitters while Feasible $TF$ \emph{in every simulation}, with the performance gain being more prominent for networks with larger $N$. Interestingly, Most-Overlaps Sort and Bandwidth-Coverage Sort show a significant decline in their ability to allocate transmitters within the available bandwidth $F$ as the number of transmitters $N$ increases, while Least-Bandwidth Sort and Least-Coverage Sort continue to benefit from a larger number of transmitters. %increased transmitter counts \( N \to \infty \),.

Intuitively, this effect can be explained by the manner in which the different sorting algorithms prioritize the increasing number of transmitters $N$. %that become available as $N$ increases. 
Least-Bandwidth Sort and Least-Coverage Sort prioritize transmitters with the lowest bandwidth requirements \( B_i \) and with the smallest coverage radius $R_i$, respectively, allowing early frequency allocations to occupy minimal resources. As a result, a large number of transmitters can be allocated while the network remains feasible, leading to a large $TF$. Latter frequency allocations have high bandwidth requirements $B_i$ and/or high coverage areas $R_i$, increasing the chances of the network becoming unfeasible, leading to a low $FI$. 
%As more transmitters become available in the network, the algorithm finds transmitters with even lower \( B_i \) than those initially allocated. 
%Similarly, \emph{Least-Coverage Sort} prioritizes transmitters with the smallest coverage radius $R_i$, which, consequently, are less likely to interfere with neighboring transmitters. %have coverage overlaps. %the least likelihood of interference, or geographic overlap, with other allocated transmitters. 
%As a result, a large number of transmitters (each with small coverage area) can be allocated while the network remains feasible, leading to a large $TF$. Latter frequency allocations have high coverage areas and (potentially) many coverage overlaps, thus increasing the chances of the network becoming unfeasible, leading to a low $FI$. 
%This translates to denser allocation opportunities within the spectrum bandwidth, hence facilitating additional \emph{N} transmitters. Similarly, as the transmitter count \( N \to \infty \), the algorithm finds transmitters with smaller radii \( R_i \) than those previously allocated. 
Most-Overlaps Sort and Bandwidth-Coverage Sort prioritize transmitters with most coverage overlaps and highest Bandwidth-Coverage Product, respectively. By starting with the most challenging transmitters, these algorithms increase the chances of finding a feasible set of frequency allocations for all $N$ transmitters, leading to a large $FI$. On the other hand, this prioritization also increases the chances of the network becoming unfeasible with only a few allocated transmitters, leading to a low $TF$. Prioritizing the most challenging transmitters leads to ``all-or-nothing'' results in which the network is either feasible (with $TF=N$) or unfeasible with low $TF$. %As the availability of transmitters expands, the algorithms allocate more spectrum-intensive transmitters, which occupy larger amounts of spectrum space for fewer transmitters. 
%Transmitters first allocated by Most-Overlaps Sort will pose greater chances of signal interference due to their larger sizes, making it difficult to exploit spectrum sharing opportunities. 
While Bandwidth-Coverage Sort prioritizes transmitters that consume most resources, it doesn't necessarily select those with the most overlaps, which have the lowest potential for spectrum sharing, explaining the relatively better performance of Bandwidth-Coverage Sort when compared to Most-Overlaps Sort.
%It is evident from Figure \ref{fig:simulation_increasing_N}(b) that network inconsistency, as seen through the error margins, is a major disadvantage of the trailing algorithms.

To further investigate the behavior of Most-Overlaps Sort, in Figure~\ref{fig:simulation_increasing_N}(d) we show the results of a homogeneous networks with $R_i = 12$ meters and $B_i = 2$ units, $\forall i$. In this homogeneous network, Least-Coverage Sort, Least-Bandwidth Sort, Bandwidth-Coverage Sort, and Random Sort are rendered equivalent, and they all outperform Most-Overlaps Sort in terms of the Total Transmitters while Feasible $TF$, suggesting that DSA algorithms that prioritize transmitters with most overlaps, or high-degrees, can be extremely inefficient. Notice that as $N$ increases, new transmitters with high-degree are prioritized by Most-Overlaps Sort, affecting the frequency allocation of transmitters with lower degrees, thereby (potentially) further decreasing $TF$.

% [Need to add an analysis of Figure 4 here] 
In Figures~\ref{fig:simulation_increasing_F}(a) and (b), we simulate networks with increasing total available bandwidth $F\in\{5,6,\ldots,15\}$ and show the performance of DSA algorithms with respect to Total Transmitters while Feasible, $TF$, and the Coverage Area, $CA$.  
In general, as the available spectrum $F$ increases, all DSA algorithms are able to allocate more admissible transmitters which, in turn, increases the coverage area. %and, therefore, increase the overall coverage area. % in the given spectrum space $F$. 
When the available spectrum is high enough $F \geq 14$, most of the $N=25$ transmitters become admissible, and the performance gaps between the different DSA algorithms is reduced. 

The results in Figure~\ref{fig:simulation_increasing_F}(a) align with Figure~\ref{fig:simulation_increasing_N}(b), showing that Least-Bandwidth Sort and Least-Coverage Sort outperform Most Overlaps Sort and Bandwidth-Coverage Sort in terms of $TF$. 
As expected, the results in Figure~\ref{fig:simulation_increasing_F}(b) show that Least-Coverage Sort under-performs in terms of coverage area $CA$. 
Least-Bandwidth Sort achieves the highest $TF$ in every simulation and the highest $CA$ in bandwidth-constrained networks with $F \leq 6$
since it prioritizes transmitters with the lowest bandwidth requirements $B_i$, thereby maximizing the number of admissible transmitters within $F$ and the corresponding coverage area. 
When $6 < F \leq 12$, the coverage area $CA$ of Least-Bandwidth Sort is comparable with the benchmark algorithm, Random Sort, and 
%
%In the mid-range of $F$, specifically when \( 6 < F \leq 12 \), 
Bandwidth-Coverage Sort leads by a significant margin. Bandwidth-Coverage Sort considers both \( B_{\text{i}} \) and \( R_{\text{i}} \), ensuring, to some extent, that transmitters with large coverage areas are prioritized, leading to high $CA$. %Since \emph{CA} is dependent on \( R_{\text{i}} \) for  $i\in\{1,2\ldots,N\}$ transmitters, prioritizing a greater \( R_{\text{i}} \) results in the allocation of transmitters with the largest Coverage Areas, leading to a higher cumulative $CA$. On the other hand, Least-Radius Sort significantly under-performs for all \( F \leq 12 \), as it focuses on transmitters with the smallest \( R_{\text{i}} \). As a result, Least-Radius Sort would only allocate transmitters with the least Coverage Areas. 

% \begin{figure}[h]
% \vspace{-4mm}
% \centering
% \centerline{\includegraphics[width=0.83\columnwidth]{images/BC_MaxRad_final.png}}
% \vspace{-2mm}
% \caption{
%     Bandwidth-Coverage Product $BC$ and Maximum Transmitter Radius $R_{max}$ for all DSA algorithms. $R_{max}$ varies independently such that  \(R_{max} \in \{8, 9, \dots, 30\} \) units while other parameters are set according to Table \ref{tab:sim_parameters}.
%    Results are averaged over 50 simulation runs.}
% \label{fig:simulation_increasing_R}
% \vspace{-2mm}
% \end{figure}
% \begin{figure}[h]
% \vspace{-2mm}
% \centering
% \centerline{\includegraphics[width=0.83\columnwidth]{images/BC_maxN_final.png}}
% \vspace{-2mm}
% \caption{
%     Bandwidth-Coverage Product $BC$ and Maximum Transmitter Bandwidth $B_{max}$ for all DSA algorithms. $B_{max}$ varies independently such that $1 \leq B_{max} \leq 8$ units while the remaining parameters are as defined in  Table \ref{tab:sim_parameters}. Results are averaged over 50 simulation runs.
%     }
% \label{fig:simulation_increasing_B}
% \vspace{-2mm}
% \end{figure}
In Figures~\ref{fig:simulation_increasing_R_B}(a) and (b), we simulate networks with increasing heterogeneity and display the performance of the DSA algorithms in terms of their Bandwidth-Coverage Product, $BC=\sum_{i\in\mathcal{N}} R_i B_i$. Recall that a high $R_i B_i$ indicates a transmitter that consumes a significant amount of resources, which may represent an important transmitter that can (potentially) serve many receivers and can support high quality of service. Hence, $BC$ can serve as an indicator of whether the DSA algorithm can allocate important transmitters. Figure~\ref{fig:simulation_increasing_R_B}(a) and (b) considers transmitters with increasing coverage radius $R_i$ and bandwidth requirements $B_i$, respectively. 
%Figure \ref{fig:simulation_increasing_R} models the Bandwidth-Coverage Product $BC$ for increasing maximum transmitter radii \(R_{max} \in \{8, 9, \dots, 30\} \) while Figure \ref{fig:simulation_increasing_B} depicts networks with increasing maximum transmitter radii \(B_{max} \in \{1, 2, \dots, 8\} \)  to examine the performance in terms of the Bandwidth-Coverage Product $BC$. 
As expected, Bandwidth-Coverage Sort outperforms other DSA algorithms in every simulation, with 
%in Figure \ref{fig:simulation_increasing_R}, Bandwidth-Coverage Sort returns a significantly greater $BC$ across the entire range of maximum transmitter radii $R_{max}$, with 
the improvement becoming more noticeable as the networks become more heterogeneous. %when \( R_{max} > 20 \). Bandwidth-Coverage Sort is inherently designed to allocate transmitters with the highest $BC$, or the ones which demand the most resources. 
Least-Bandwidth Sort and Least-Coverage Sort attain the lowest $BC$, consistently performing below the benchmark algorithm. %Least-Bandwidth Sort selects transmitters with the lowest bandwidth consumption, which is one component of the $BC$ calculation. Similarly, Least-Coverage Sort aims to minimize the coverage area at the start of allocation, which directly affects the other component of the $BC$ value. In essence, both these DSA algorithms prioritize minimizing their respective resource consumptions—bandwidth for Least-Bandwidth Sort and coverage area for Least-Coverage Sort—leading to a significant reduction in the overall $BC$. As more variation in the maximum transmitter radius $R_{max}$ becomes possible, as $R_{max}$ increases, the difference becomes more pronounced. 
%
%Figure \ref{fig:simulation_increasing_B} also displays Bandwidth-Coverage Sort attaining superior levels of $BC$ due to its direct correlations with transmitter bandwidth and coverage. 
Interestingly, Most-Overlaps Sort also under-performs in terms of $BC$, with a performance that is either close to or below Random Sort, as can be seen in Figure~\ref{fig:simulation_increasing_R_B}(b) when \(B_{max} > 4 \). %Least-Bandwidth Sort, and Least-Coverage Sort all fail to maintain a higher $BC$ than the benchmark algorithm Random Sort when \(B_{max} > 4 \), 
These results suggest that neither Least-Bandwidth Sort, Least-Coverage Sort, or Most-Overlaps Sort are good options for facilitating important transmitters which support a higher quality of service. 

Figure~\ref{fig:simulation_increasing_R_B}(b) shows that the DSA algorithms exhibit a sharp decrease in $BC$ when \(B_{\text{max}} > 7\). This decline arises from the increased likelihood of encountering transmitters with demanding bandwidth requirements. %As \(B_{\text{max}}\) increases, the probability of having transmitters with large bandwidth demands grows, which constrains spectrum sharing. 
These demanding transmitters limit the total number of transmitters that may be allocated feasibly $TF$, indirectly reducing the overall $BC$. %While some transmitters may exhibit higher individual $BC$ values as $B_{max}$  grows, the decreased number of allocated transmitters $N_{allocated}$ ultimately results in a lower total $BC$. 
Notice that Bandwidth-Coverage Sort, which prioritizes these demanding transmitters, experiences this decline earlier than other DSA algorithms. %as it decreases as soon as \(B_{\text{max}} > 5\)

\vspace{0.5ex}
\noindent\textbf{Summary of Numerical Results.} Figures~\ref{fig:simulation_increasing_N},~\ref{fig:simulation_increasing_F}, and~\ref{fig:simulation_increasing_R_B} compare different DSA algorithms in terms of feasibility, $FI$, number of transmitters allocated before the network becomes unfeasible, $TF$, and resource utilization, $BU$, $BC$, and $CA$. Least-Coverage Sort and Least-Bandwidth Sort maximize the number of transmitters allocated before the network becomes unfeasible, having a comparable performance in most simulations. As expected, Least-Bandwidth Sort outperforms Least-Coverage Sort in terms of coverage area $CA$. Most-Overlaps Sort and Bandwidth-Coverage Sort maximize the chances of feasibility. Bandwidth-Coverage Sort outperforms Most-Overlaps Sort -- which is a popular algorithm in the literature -- in the vast majority of the simulations.

%% file: src/conclusion.tex
% This paper presented a novel mechanism for performing spectrum coordination using SCMs which offer a standardized means to capture the spectral, temporal, and spatial characteristics of spectrum use of RF devices and systems. The proposed algorithm dynamically adjusts the transmission parameters, ensuring aggregate compatibility among all existing RF devices. The algorithm was evaluated in terms of computation time, efficiency of spectrum allocation, and number of device reconfigurations using a custom platform that simulates dynamic and
% dense communication environments. The simulation results in this paper and the experimental validation in~\cite{stojadinovicspectrum} demonstrate the feasibility of our SCM-based spectrum deconfliction technique in performing fine grained spectrum assignments at scale. 

% In future work, we will consider networks with mmWave nodes using phased array antennas and develop enhanced SCM-based DSA algorithms that take into account antenna directionality and develop machine learning-based algorithms to enable efficient SCM based spectrum access schemes.

In this paper, we considered heterogeneous 6G networks with $N$ base stations (called transmitters) and a total of $F$ frequency bands available for transmission. Each transmitter is associated with a geographical location, a coverage radius $R_i$, and a bandwidth requirement $B_i$. We used extensive simulation results to evaluate the performance of different DSA algorithms in various heterogeneous networks. %in terms of the feasibility of the frequency allocation task $FI$, the number of transmitters allocated before the network becomes unfeasible $TF$, and resource utilization, namely bandwidth usage $BU$, coverage area $CA$, and bandwidth-coverage product $BC$. 
Based on the results, we separate the DSA algorithms in two classes: 
(i) Most-Overlaps Sort and Bandwidth-Coverage Sort, which prioritize the most challenging (i.e., resource hungry) transmitters; and 
(ii) Least-Coverage Sort and Least-Bandwidth Sort, which prioritize transmitters that use minimal resources.
Class (i) leads to ``all-or-nothing'' results. They achieve the highest values of a feasible frequency allocation ($FI$) for all $N$ transmitters, which can be seen as the ultimate objective of DSA. However, when the network is unfeasible, these algorithms tend to significantly under-perform when compared with Class (ii), especially in terms of the number of transmitters, $TF$, allocated while the network was feasible. %In contrast, Class (ii) maximizes the number of transmitters allocated while the network was feasible. 
%This paper evaluated DSA-based sorting approaches designed to optimize spectrum allocation in 6G wireless networks. Our network model featured multiple transmitters, each characterized by varying coverage areas, bandwidth requirements, and geographical distributions. The challenge involved allocating spectrum resources systematically in order to mitigate interference and maximize the efficiency of spectrum utilization with respect to feasibility and transmitter assigning capacity. The simulation results in this paper demonstrate that service providers (SPs) with a focus on maintaining the overall feasibility of their networks (servicing all their customers) may benefit from using Bandwidth-Coverage Sort and Most-Overlaps Sort. While high degree allocation performance has been well discussed in literature, the low degree sorts were generally the most effective algorithms in scenarios where the target metric was the number of transmitters (Fig. 5), suggesting that SPs aiming to facilitate the greatest number of users within the available spectrum should specifically avoid Most-Overlaps Sort and consider employing Least-Coverage Sort or Least-Bandwidth Sort–algorithms that minimize resource consumption. 
%
Interesting extensions include consideration of base stations with directional antennas, time-varying bandwidth requirements and that can enter/leave the network. 
%In future work, we look to refine and test our proposed algorithms under dynamic conditions where transmitters continuously enter and exit the geographic space, aiming to evaluate sustained performance over extended periods. Additionally, we will consider implementing machine learning techniques to predict network demand in real-time (comparing them with our algorithmic approaches) and assist with DSA-based efforts in spectrum optimization. 

%We utilize a modified version of the interaction language CIL developed by DARPA to enable the exchange of SCM messages between wireless networks
%where SCM based spectrum use deconfliction  was modeled